\documentclass[prd,twocolumn,amsmath,amssymb,axodraw]{revtex4}
\usepackage{graphicx}
\def\Journal#1#2#3#4{{#1} {\bf #2}, #3 (#4)}

\def\PLB{{\em Phys. Lett.}  B}

\def\be{\begin{equation}}
\def\ee{\end{equation}}
\def\bea{\begin{eqnarray}}
\def\eea{\end{eqnarray}}

\begin{document}
\vspace*{4cm}
\title{RECENT BES RESULTS ON $\psi(3770)$ AND $D$ MESON PRODUCTION AND
DECAY\footnote{The talk given at 39th Recontres de Moriond on 
Electroweak Interaction and Unified Theories, LaThuile, Aosta Valley,
Italy, 21-28 March, 2004.}}

\author{ G. RONG \\ (for the BES Collaboration)}

\address{Institute of High Energy Physics, Beijing, 100039, China}

\begin{abstract}
Using a data sample of $17.7$ $\rm pb^{-1}$ collected at 3.773 GeV
with  the BES-II detector at the BEPC,
the cross sections for
$D^0 \overline D^{0}$ and $D^+D^-$ productions at 3.773 GeV have
been measured.
From the data sample about 33 $\rm pb^{-1}$ taken around 3.773 GeV,
$7696 \pm 199 \pm 369$ and
$5381 \pm 128 \pm 188$ singly-tagged neutral and charged $D$
mesons are accumulated, respectively. In the system recoiling against
the singly-tagged charged $D$ sample, 3 purely leptonic decay events
of $D^+ \rightarrow \mu^+\nu$ are observed, which yields the branching
fraction to be 
$BF(D^+ \rightarrow \mu^+\nu)=(0.120^{+0.092+0.010}_{-0.063-0.009}) \%$ 
and the decay constant $f_D=(365^{+121+32}_{-113-28})$ MeV. 
From the singly tagged $D$ sample, the
semileptonic decay branching fractions for $D^0 \rightarrow K^-e^+\nu$,
$D^0 \rightarrow \pi^-e^+\nu$ and 
$D^+ \rightarrow \overline K^0 e^+\nu$ are measured to be
$BF(D^0 \rightarrow K^-e^+\nu_e)=(3.52 \pm 0.36\pm 0.25)\%$,
$BF(D^0 \rightarrow \pi^-e^+\nu_e)=(0.36 \pm 0.14\pm 0.03)\%$ and
$BF(D^+ \rightarrow \overline K^0 e^+\nu_e)=(8.64 \pm 1.51\pm 0.72)\%$. 
The vector form factors are determined to be
$|f^K_+(0)| = 0.75 \pm 0.04 \pm 0.03$ and
$|f^{\pi}_+(0)| = 0.76 \pm 0.15 \pm 0.06$. 
The ratio of the two form factors is extracted to be
$|f^{\pi}_+(0)/f^K_+(0)|= 1.01 \pm 0.20 \pm 0.08$.
The ratio of the decay widths is measured to be 
$\Gamma(D^0 \rightarrow K^-e^+\nu)/\Gamma(D^+ \rightarrow \overline K^0 e^+\nu)
=1.04\pm 0.21 \pm 0.08$.
From the data sample of about 27 $\rm pb^{-1}$, the evidence of 
$\psi(3770) \rightarrow J/\psi \pi^+\pi^-$ non-${D \overline D}$ decay
is observed. The branching fraction is determined to be 
$BF(\psi(3770) \rightarrow J/\psi \pi^+\pi^-)=(0.34\pm 0.14 \pm 0.08)\%$, 
corresponding to the partial width of 
$\Gamma(\psi(3770) \rightarrow J/\psi \pi^+\pi^-) = (80 \pm 32 \pm 21)$ keV.
\end{abstract}

\maketitle

\section{Introduction}

At the center-of-mass energy of 3.773 GeV,
the $\psi(3770)$ resonance is produced in $e^+e^-$ annihilation,
and the open charm pairs of $D^0 \overline {D^0}$ and $D^+D^-$
are mainly produced from $\psi(3770)$ decays.
Taking the advantage of the $D\overline D$ production, 
we can use the single tag method to measure the cross sections for
$D^0\overline D^0$, $D^+D^-$ and $D \overline D$ productions
at the energy of  3.773 GeV;
using the double tag method we can measure some absolute 
decay branching fractions of neutral and charged $D$ mesons.

In this paper, we report some preliminary results 
on measurement of the cross section for 
$D \overline D$ production, measurement of the decay constant $f_D$,
measurements of the form factors $f^{\pi}_+(0)$ 
and $f^K_+(0)$ in the semileptonic decays of $D^0$ meson, measurement of
the ratio of partial width 
$\frac{\Gamma(D^0 \rightarrow K^- e^+\nu)}{
  \Gamma(D^+ \rightarrow \overline K^0 e^+\nu)}$,
and the evidence of $\psi(3770)$ decay to a non-${D \overline D}$ final
state.
\section{Cross section for $D \overline D$ production}

The data used in this analysis were collected
at the center-of-mass energy of 3.773 GeV
with the Beijing Spectrometer~\cite{BES-II}
at the Beijing Electron Positron Collider.
The total integrated luminosity of the data
is  $17.7$ $\rm pb^{-1}$.
The measurements of the cross sections for
the $D^0 \overline {D^0}$, $D^+D^-$ and $D\overline D$ 
productions are made based on the analysis
of singly tagged $D^0$ and $D^+$ samples.
At center-of-mass energy $\sqrt{s}=3.773$ GeV, the
$D^0$ (Through this paper, charge conjugation is implied.) and $D^+$
are produced in pair via the process of
\begin{equation}
e^+e^- \rightarrow D^0 \overline {D^0}, D^+D^-.
\end{equation}
The totally observed number $N_{D^0_{tag}}$ ($N_{D^+_{tag}}$)
of $D^0$ ($D^+$) mesons and the observed cross sections
$\sigma^{obs}_{D^0 \overline D^0}$
($\sigma^{obs}_{D^+D^-}$) are related by a relation
\begin{equation}
 \sigma_{D^0\overline D^0}^{obs} = \frac {N_{D^0_{tag}}}
                       {2 \times L \times B \times \epsilon },
\end{equation}
\noindent
and
\begin{equation}
 \sigma_{D^+ D^-}^{obs} = \frac {N_{D^+_{tag}}}
                       {2 \times L \times B \times \epsilon },
\end{equation}

\noindent
where $L$ is the integrated luminosity of the data set used in the
analysis, $B$ is the well-known branching fractions~\cite{PDG02}
for decay modes in question,
and  $\epsilon$ is the Monte Carlo efficiency for reconstruction
of this decay mode in the data analysis.

\begin{figure}[hbt]
\includegraphics[width=8.0cm,height=8.5cm]{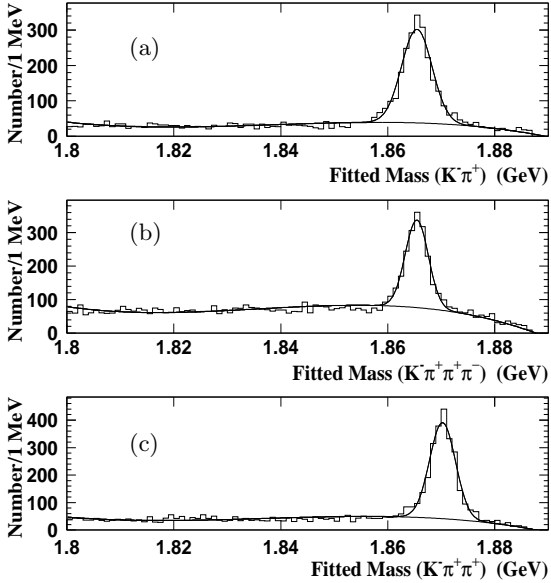}
\put(-170.0,210.0){(a)}
\put(-170.0,140.0){(b)}
\put(-170.0, 60.0){(c)}
\caption{Distribution of the fitted masses of the $Km\pi$ (m=1,
or 2, or 3) combinations for three singly tagged modes,
where Fig. (a) and Fig. (b) are for the decay
modes of $D^0\rightarrow K^-\pi^+$ and
$D^0\rightarrow K^-\pi^+\pi^+\pi^-$, respectively,
and the Fig. (c) is for the decay mode of
$D^+\rightarrow K^-\pi^+\pi^+$.
}
\label{stagxsct}
\end{figure}

In the measurements of the cross sections, the singly tagged neutral
and charged $D$ meson are made by fully reconstructing one $D$ meson
from the $D^0 \overline D^0$ ($D^+D^-$) pair in the invariant mass spectra
of the daughter particles from the $D$ decay.
Taking the advantage of the $D\overline D$ pair production,
we use kinematic fit to some specific particle combinations
to improve the ratio of signal to noise and mass resolution
in the invariant mass spectrum.
The distributions in the fitted masses of $Km\pi$
($m=1$, or $2$, or $3$) combinations,
which are calculated using the fitted momentum vectors
from the kinematic fit, are shown in Fig.~\ref{stagxsct}.
The signals for neutral and charged $D$ mesons production
are clearly observed in the fitted mass spectra as shown in
the Fig.~\ref{stagxsct}.

We measure the observed cross sections for
$D^+D^-$, $D^0 \overline{D^0}$ and $D \overline D$ productions to be 
     $$ \sigma_{D^+  {D^-} }^{obs} = (2.52 \pm 0.07 \pm0.24)~~ {\rm nb},$$
\noindent
     $$ \sigma_{D^0 \bar {D^0} }^{obs} = (3.26 \pm 0.09 \pm 0.25)~~ {\rm nb}.$$
\noindent
  $$ \sigma_{D \bar {D} }^{obs} = (5.78 \pm 0.11 \pm 0.45)~~ {\rm nb}.$$
After correcting to the effect of Initial State Radiation and the vacuum
polarization, we obtain the tree level cross sections to be,
    $$ \sigma_{D^+  {D^-} }^{tree} = (3.23 \pm 0.09 \pm0.34)~~ {\rm nb},$$
\noindent
     $$ \sigma_{D^0 \bar {D^0} }^{tree} = (4.18 \pm 0.12 \pm 0.37)~~ {\rm
nb}.$$
  $$ \sigma_{D \bar {D} }^{tree} = (7.42 \pm 0.14 \pm 0.67)~~ {\rm nb}.$$

\section{Purely Leptonic Decays of
$D^+ \rightarrow \mu^+\nu$ and Decay Constant $f_D$}

\begin{figure}[hbt]
\includegraphics[width=9.0cm,height=9.5cm]
{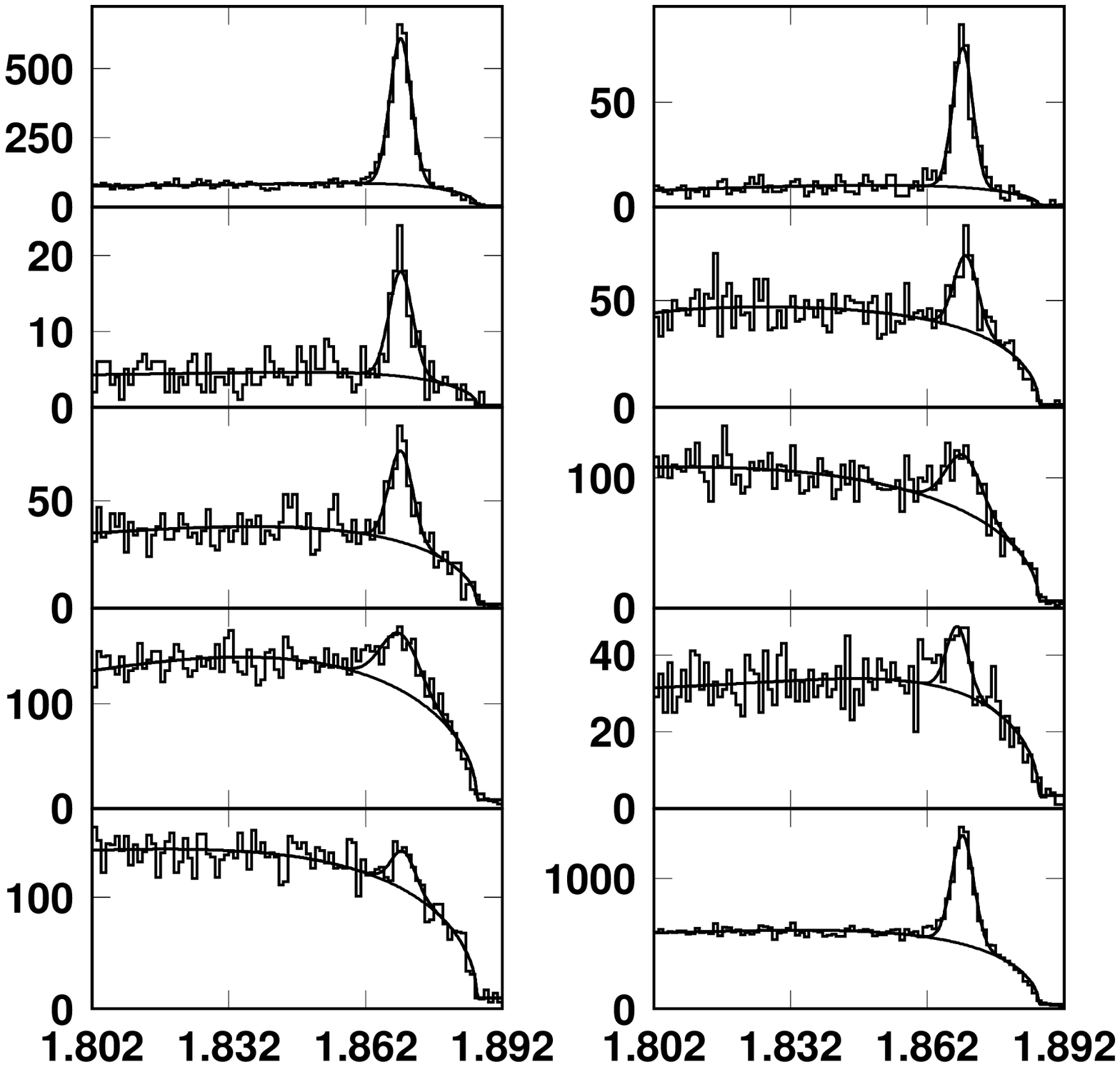}
\put(-200,240){(a)}
\put(-200,200){(c)}
\put(-200,155){(e)}
\put(-200,95){(g)}
\put(-200, 50){(i)}
\put(-90,240){(b)}
\put(-90,200){(d)}
\put(-90,135){(f)}
\put(-90,90){(h)}
\put(-90, 60){(j)}
\put(-170,-5){Invariant Mass (GeV/$c^2)$}
\put(-250,95){\rotatebox{90}{Events/(0.001 GeV/$c^2$)}}
\caption{Distribution of the fitted masses of the $nKm\pi$ (m=0 or 1 or 2,
n=1 or 2 or 3 or 4) combinations for the 9 single tag modes; (a), (b), (c), (d),
(e), (f), (g), (h) and (i) are for the modes of 
$D^- \rightarrow K^+\pi^-\pi^-$,
$D^- \rightarrow K^0\pi^-$,
$D^- \rightarrow K^0K^-$,
$D^- \rightarrow K^+K^-\pi^-$,
$D^- \rightarrow K^0\pi^-\pi^-\pi^+$,
$D^- \rightarrow K^0\pi^-\pi^0$, 
$D^- \rightarrow K^+\pi^-\pi^-\pi^0$,
$D^- \rightarrow K^+\pi^+\pi^-\pi^-\pi^-$
and $D^- \rightarrow \pi^-\pi^-\pi^+$ respectively;
(j) is the fitted masses of the $nKm\pi$
combinations for the 9 modes together.}
\label{dpsgtag}
\end{figure}

The $D^-$ mesons are reconstructed in
$K^+\pi^-\pi^-$,  
$K^0 \pi^-$,
$K^0 K^-$,
$K^+K^-\pi^-$, 
$K^0 \pi^-\pi^-\pi^+$,
$K^0 \pi^-\pi^0$,
$K^+\pi^-\pi^-\pi^0$,
$K^+\pi^+\pi^-\pi^-\pi^-$ and
$\pi^-\pi^-\pi^+$ modes
with sub-resonances decaying as
$K^0_S\rightarrow \pi^+\pi^-$, and $\pi^0 \rightarrow \gamma~\gamma$.
Fig.~\ref{dpsgtag} shows the distributions 
in the fitted masses of $mKn\pi$ ($m=1,2$ and the $n=1,2,3$) combinations.
The signals for $D^-$ production are clearly observed
in the fitted mass spectra as shown in
the Fig.~\ref{dpsgtag}.
We accumulate $5381 \pm 128 \pm 188$ singly tagged $D^-$ mesons
in total.

The system recoiling against the singly tagged $D^-$ in each of the
events as shown in the Fig. 2 are  examined for consistency with the decay
$D^+ \rightarrow \mu^+ \nu$. It is required that
there 
is a singly charged track
and $|\cos \theta| < 0.68$,
where the $\theta$ is the polar angle of the charged track.
There must be hits in the muon counter which are well associate
within $\pm 4 \sigma$ with the
tracks in transverse projection and Z direction;
the required number of hits is momentum
dependent. 

For the candidate events, no isolated photon
is allowed to be present, where the
isolated photon is defined as an electromagnetic shower with the energy
being greater than 100 MeV and the direction of the shower development
separated by at least 20 degrees from the direction
of the nearest charged track.
Since there is a missing neutrino
in the purely leptonic decay event, the event should
be characteristic with missing energy $E_{miss}$ and
missing momentum $P_{miss}$
carried by the neutrino.
For the typical purely leptonic decay events, 
the difference between the $E_{miss}$ and
the $P_{miss}$ should be around zero.
We define the deference to be
$$ U_{miss} = E_{miss} - P_{miss}. $$
To select the purely
leptonic decay events from the singly tagged $D^-$ sample,
we require the $U_{miss}$ of the candidate events to be
in the $\pm 3.0\sigma_{U_{miss}}$ region.
We also require the momentum of the muon
to be within the range of 0.780 GeV/$c$ to 1.115 GeV/$c$.
Three candidate events satisfy the selection criteria of
the purely leptonic decay events.
A detailed Monte Carlo study shows that there are $0.31\pm 0.16$
background events in the 3 candidate events. 
The efficiency to reconstruct the purely leptonic decay events
is $0.405 \pm 0.011$. These result the purely leptonic branching fraction
to be 
$$BF(D^+ \rightarrow \mu^+\nu)=(0.120^{+0.092+0.010}_{-0.063-0.009})\%,$$
\noindent
and decay constant of 
$$f_D = (365^{+121+32}_{-113-28})~\rm MeV,$$
\noindent
where the first errors are statistical and second systematic which arise
from the uncertainties in the measured branching fraction, the CKM
matrix element $|V_{cd}|$ and the lifetime of $D^+$; the
systematic uncertainties in the measured branching fraction 
arise mainly from the uncertainties in the $\mu^+$ identification, 
tracking efficiency, background subtraction and $U_{miss}$ cut.
The central value of the $f_D$ is consistent with that measured
with BES-I detector~\cite{bes1fd}.

\section{Semileptonic Decays of Charged and Neutral $D$ mesons}

\subsection{Branching fractions for $D^0 \rightarrow K^-e^+\nu$ and
$D^0 \rightarrow \pi^-e^+\nu$
}

\begin{figure}[hbt]
\includegraphics[width=8.5cm,height=8.5cm]
{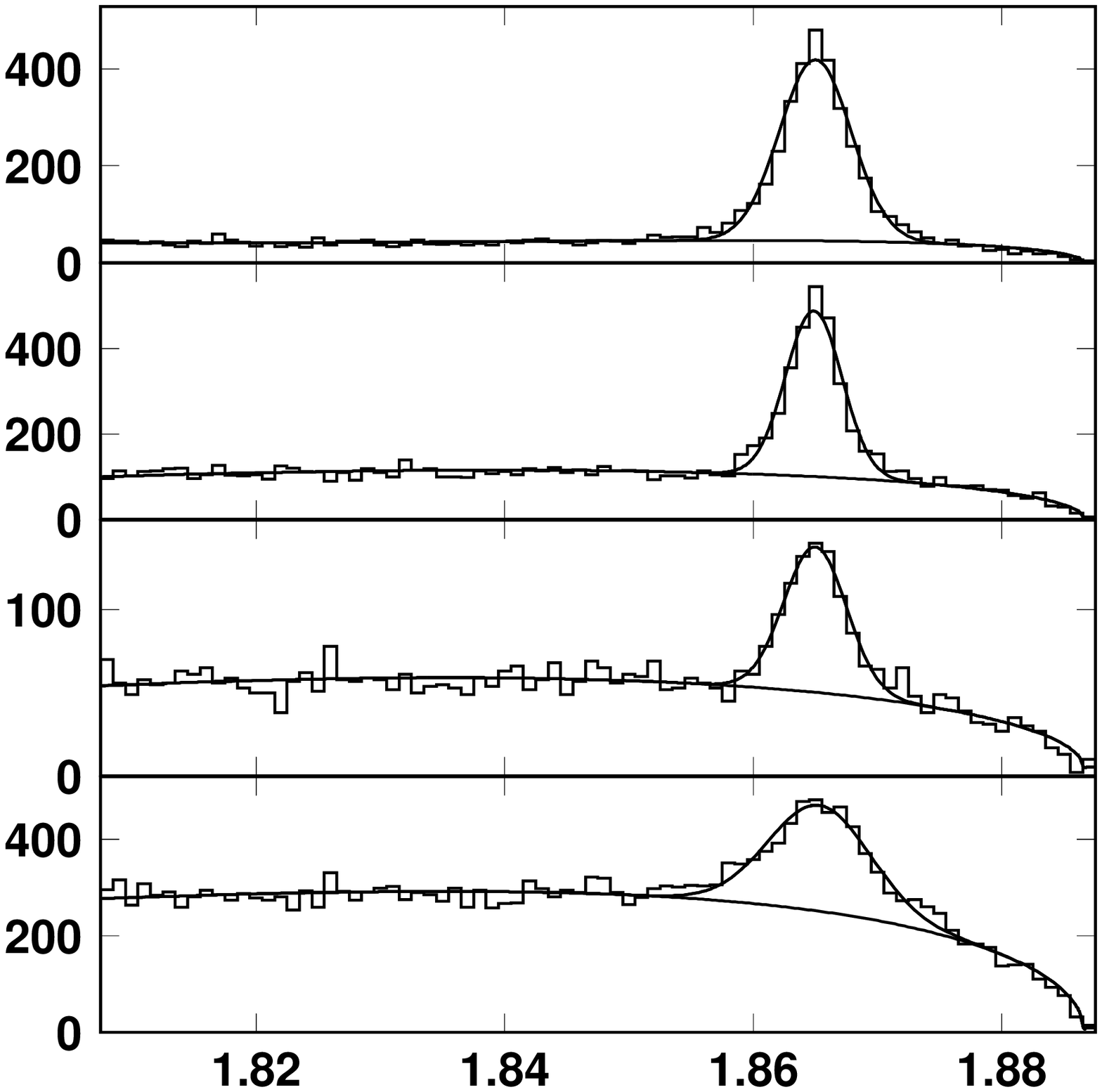}
\put(-190,205){(a)}
\put(-190,160){(b)}
\put(-190,110){(c)}
\put(-190,65){(d)}
\put(-150,-5){Invariant Mass (GeV/$c^2)$}
\put(-245,85){\rotatebox{90}{Events/(0.001 GeV/$c^2$)}}
\caption{Distributions of the fitted invariant masses of (a) $ K^+\pi^-$,
(b) $K^+\pi^-\pi^+\pi^-$, (c) $ K^0_S\pi^+\pi^-$ and
(d) $ K^+\pi^-\pi^0$ combinations.}
\label{sgtag}
\end{figure}

\begin{figure}[hbt]
\includegraphics[width=8.5cm,height=8.5cm]
{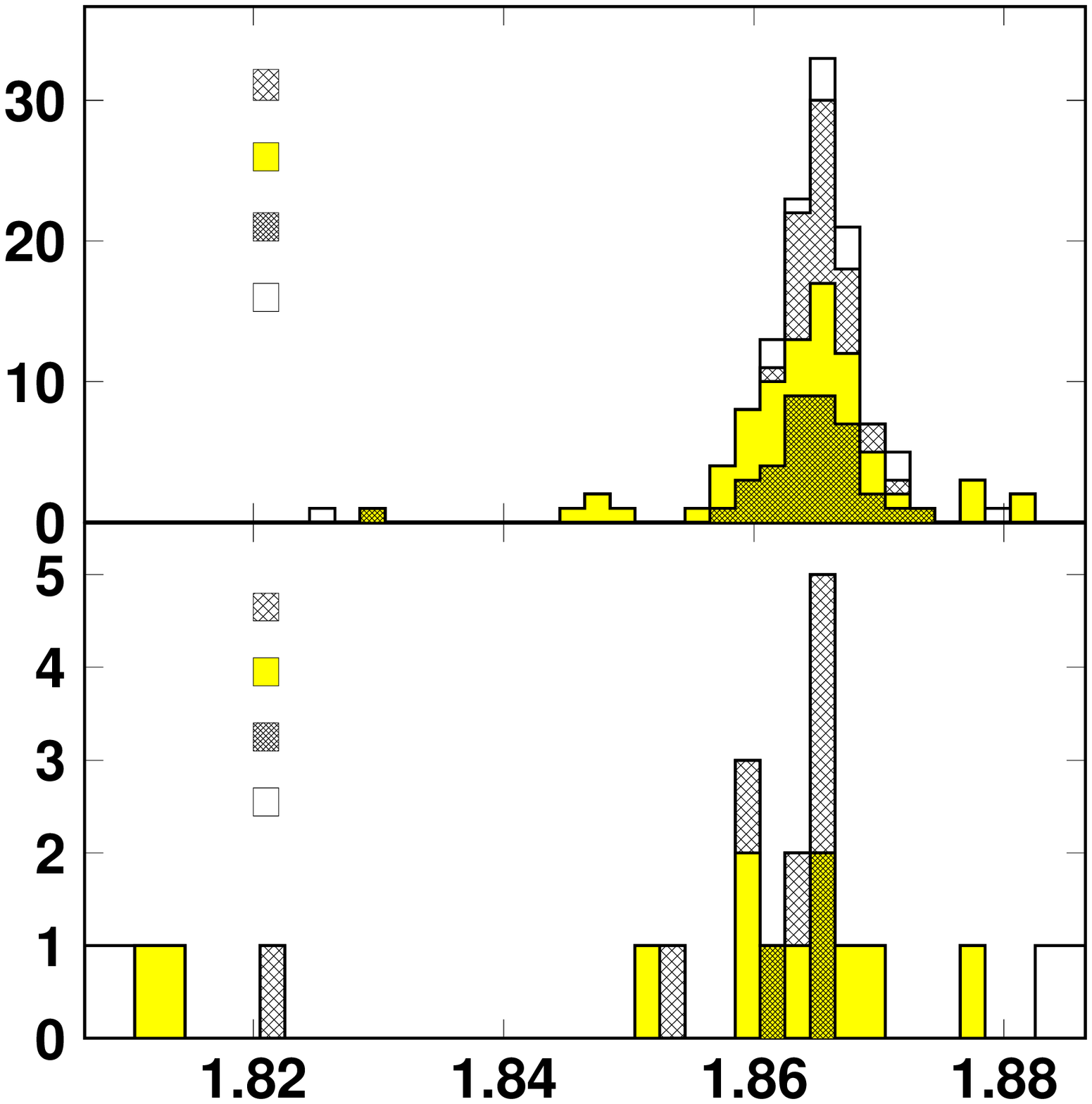}
\put(-190,200){\bf{(a)}}
\put(-190,115){\bf{(b)}}
\put(-150,-7){Invariant Mass (GeV/$c^2)$}
\put(-245,70){\rotatebox{90}{Events/(0.002 GeV/$c^2$)}}
\caption{Distributions of the fitted invariant masses of $Kn\pi$
combinations for the events for which (a) the
          $D^0 \rightarrow K^-e^+\nu_e$ and
          (b) the $D^0 \rightarrow \pi^-e^+\nu_e$
          candidate events are observed in the system recoiling against the
tagged $\overline D^0$.}
\label{d0tokpiev}
\end{figure}

The singly tagged $\overline D^0$ mesons are obtained by reconstructing
the four decay modes of $\overline D^0 \rightarrow K^+\pi^-$,
$\overline D^0  \rightarrow K^+\pi^-\pi^-\pi^+$,
$\overline D^0 \rightarrow K^0\pi^+\pi^-$
and
$\overline D^0 \rightarrow K^+\pi^-\pi^0$.
Fig.~\ref{sgtag} shows the fitted mass distributions
of $Kn\pi$ ($n=1,2,3$)
combinations. We totally accumulate $7696 \pm 199 \pm 369$
singly tagged neutral $D$ meson sample.
Fig.~\ref{d0tokpiev}(a) and Fig.~\ref{d0tokpiev}(b)
show the distribution of
the fitted masses of the $Kn\pi$ combinations
for the events for which the  $D^0 \to K^-e^+\nu_e$ and
$D^0 \to \pi^-e^+\nu_e$ candidate events are observed
in the system recoiling against the singly tagged $\overline D^0$.
After subtracting numbers of background events
(sideband background events and the background events
due to other semileptonic or hadronic decays),
$100.5\pm 10.2$ and $10.3 \pm 3.9$ signal events
for $D^0 \rightarrow K^-e^+\nu _e$ and
$D^0 \rightarrow \pi ^-e^+\nu _e$ decays are retained.
We measure the branching fractions to be
$$BF(D^0 \to K^-e^+\nu_e)=(3.52\pm 0.36 \pm 0.25)\%,$$
$$BF(D^0 \to \pi^- e^+\nu_e)=(0.36\pm 0.14\pm 0.03)\%.$$

\subsection{Branching fraction for $D^+ \rightarrow \overline K^0e^+\nu$}

Candidate events $D^+\rightarrow \overline K^0e^+\nu _e$
is selected from the surviving tracks in the system recoiling against the
tagged $D^-$. To select the $D^+\rightarrow \overline K^0e^+\nu _e$
events, it is required that there are only
three charged tracks, one of which is identified as an electron and
the other two tracks as $\pi^+$ and $\pi^-$.
The candidate events are required to satisfy
the requirement $|U_{miss}| < 3\sigma_{U_{miss}}$, where the
$\sigma_{U_{miss}}$ is
the standard deviation of the $U_{miss}$ distribution.

\begin{figure}[hbt]
\includegraphics[width=9.0cm,height=7.0cm]
{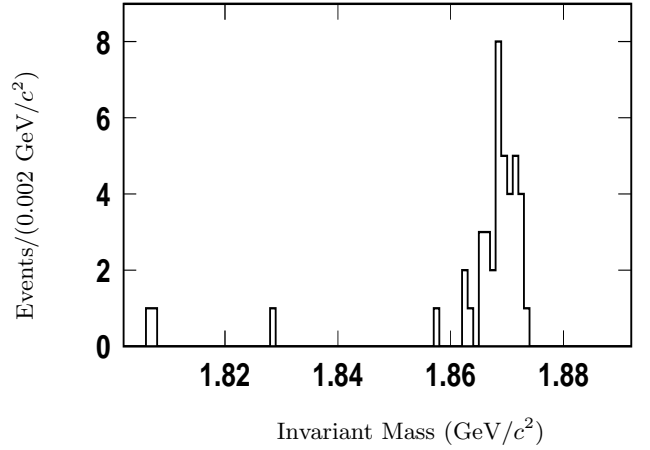}
\put(-150,15){Invariant Mass (GeV/$c^2)$}
\put(-250,60){\rotatebox{90}{Events/(0.002 GeV/$c^2$)}}
\caption{Distributions of the fitted masses of $Kn\pi$
combinations for the events for which the
          $D^+ \rightarrow \overline K^0 e^+\nu_e$
          candidates are observed in the system recoiling against the
tagged $D^-$.}
\label{dptok0bev}
\end{figure}

Fig.~\ref{dptok0bev}
show the distribution of the fitted invariant masses of the $Kn\pi$
combinations
for the events for which the  $D^+ \rightarrow \overline K^0 e^+\nu_e$
candidate events are observed in the system recoiling against
the singly tagged $D^-$. We measure the branching fraction to be
$$BF(D^+ \to \overline K^0 e^+\nu_e)=(8.64\pm 1.51 \pm 0.72)\%.$$

\subsection{Form factors $f^{\pi}_+(0)$ and $f^{K}_+(0)$}
The relations~\cite{morrison}~\cite{medina} 
between the decay widths and the form factors are,
\begin{equation}
\Gamma(D^0 \to K^-e^+\nu_e) = 1.53\;|V_{cs}|^2
|f^K_+(0)|^2 \times 10^{11} s^{-1},
\label{eq:gamma_kev}
\end{equation}
\begin{equation}
\Gamma(D^0 \to \pi^-e^+\nu_e) = 3.01\; |V_{cd}|^2
|f^{\pi}_+(0)|^2\times 10^{11} s^{-1}.
\label{eq:gamma_piev}
\end{equation}
The form factors $|f^K_+(0)|$ and $|f^{\pi}_+(0)|$
can be extracted by using the measured values
of the branching fractions and the lifetime
of the $D^0$ meson. Inserting the values of $|V_{cs}| = 0.996 \pm 0.013$,
$|V_{cd}| = 0.224 \pm 0.016$ and the lifetime
$\tau_D^0 = (411.7 \pm 2.7) \times 10^{-15}$~\cite{PDG02} into equation
(\ref{eq:gamma_kev}) and (\ref{eq:gamma_piev}), the form factors are
extracted to be
$|f^K_+(0)| = 0.75 \pm 0.04 \pm 0.03$,
$|f^{\pi}_+(0)| = 0.76 \pm 0.15 \pm 0.06,$
\noindent
where the first errors are statistical and the
second are systematic.
We extract the ratio of the two form factors to be
$$|f^{\pi}_+(0)/f^K_+(0)|= 1.01 \pm 0.20 \pm 0.08.$$

\subsection{The ratio of $\Gamma(D^0 \rightarrow K^- e^+\nu)/
                          \Gamma(D^+ \rightarrow \overline K^0 e^+\nu)$}

Isospin symmetry predicts that the ratio of partial widths
$\Gamma(D^0 \rightarrow K^- e^+\nu)/
 \Gamma(D^+ \rightarrow \overline K^0 e^+\nu)$
should be unity.
The ratio can be obtained from the
measured branching fractions for $D^0 \rightarrow K^- e^+\nu$
and $D^+ \rightarrow \overline K^0 e^+\nu$, and the lifetimes
of the $D^0$ and $D^+$ mesons.
\begin{equation}
\frac{\Gamma(D^0 \rightarrow K^- e^+\nu)}{
  \Gamma(D^+ \rightarrow \overline K^0 e^+\nu)}=
  \frac{BF(D^0 \rightarrow K^- e^+\nu)}
       {BF(D^+ \rightarrow \overline K^0 e^+\nu)}
\frac{\tau_{D^+}}{\tau_{D^0}},
\label{eq:ratwidth}
\end{equation}
\noindent
where the $\tau_{D^+}$ and $\tau_{D^0}$ are the lifetimes
of $D^+$ and $D^0$ mesons.
Inserting the measured branching fractions, and the lifetimes of the $D^+$ and
$D^0$ mesons, we obtained the ratio to be
$$\frac{\Gamma(D^0 \rightarrow K^- e^+\nu)}{
  \Gamma(D^+ \rightarrow \overline K^0 e^+\nu)}=1.04\pm0.21\pm 0.08,$$
\noindent
which is consistent with that the isospin conservation is held
in $D$ meson decays. 

\section{Evidence of $\psi(3770) \rightarrow J/\psi \pi^+\pi^-$}
The $\psi(3770)$ resonance is believed to be a mixture of the
$1 ^3D_1$ and $2 ^3S_1$ states of
the $c \bar c$ system~\cite{mark1}.  Since its mass is above open
charm-pair threshold and its width is two orders of magnitude larger
than that of the $\psi(2S)$,
it is thought to decay almost entirely to pure $D \overline
D$~\cite{Bacino}.  However, recently
some theoretical calculations point out that the $\psi(3770)$ could
decay to non$-D \overline D$ final states~\cite{Lipkin}~\cite{kuangyp}.

To search for the decay of $\psi(3770) \rightarrow J/\psi \pi^+\pi^-$,
$J/\psi \rightarrow e^+e^-$ or $\mu^+\mu^-$, $\mu^+\mu^- \pi^+ \pi^-$ and
$e^+e^- \pi^+ \pi^-$ candidate events are selected.
These are required to have four charged tracks with
zero total charge.
Candidate events are subjected to four-constraint kinematic fits to
either the $e^+e^- \rightarrow \mu^+\mu^- \pi^+ \pi^-$ or the $e^+e^-
\rightarrow e^+e^- \pi^+ \pi^-$ hypothesis.

\begin{figure}[hbt]
\includegraphics[width=8.5cm,height=11.0cm]{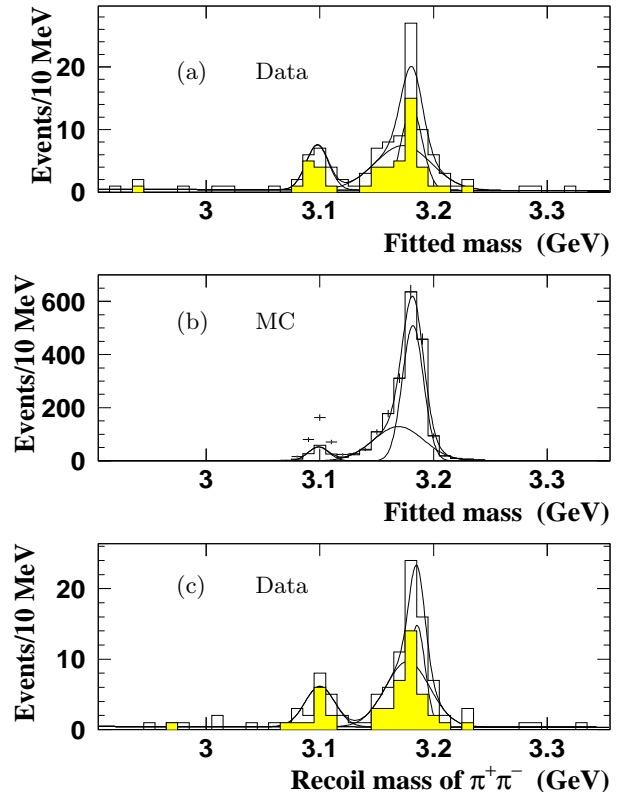}
\put(-180,270){{(a)}}
\put(-150,270){{Data}}
\put(-180,175){{(b)}}
\put(-150,175){{MC}}
\put(-180,75){{(c)}}
\put(-150,75){{Data}}
\caption{Distributions of dilepton masses for (a) data and (b) Monte
Carlo sample. The hatched histogram in (a) is for $J/\psi \rightarrow
\mu^+\mu^-$, while the open one is for $J/\psi \rightarrow
e^+e^-$. The histogram in (b) is for $\psi(2S)\rightarrow J/\psi
\pi^+\pi^-$, while the points with error bars are the sum of
$\psi(3770)\rightarrow J/\psi \pi^+\pi^-$ and $\psi(2S)\rightarrow
J/\psi \pi^+\pi^-$.  (c) Distribution of mass recoiling against the
$\pi^+\pi^-$ system calculated using measured momenta for events that
pass the kinematic fit requirement, where the hatched histogram
is for $J/\psi \rightarrow \mu^+\mu^-$ and the open one is for
$J/\psi \rightarrow e^+e^-$.}
\label{fitmassplot}
\end{figure}

Fig.~\ref{fitmassplot}(a) shows the dilepton
masses determined from the fitted lepton momenta of the accepted
events.  There are clearly two peaks. The lower mass peak is mostly
due to $\psi(3770) \rightarrow J/\psi \pi^+\pi^-$, while the higher
one is due to $\psi(2S) \rightarrow J/\psi \pi^+\pi^-$.  Since the
higher peak is produced by radiative return to the $\psi(2S)$ peak,
its energy will be approximately 3.686 GeV, while the c.m. energy is
set to the nominal energy in the kinematic fitting.  Therefore,
the dilepton masses calculated based on the fitted lepton momenta from
$\psi(2S) \rightarrow J/\psi\pi^+\pi^-$, $J/\psi \rightarrow l^+ l^-$ 
are shifted upward to about 3.18 GeV.
The fit to this peak
yields a $J/\psi$ signal of $17.8\pm 4.8$
events.

There is a contribution from $\psi(2S) \rightarrow J/\psi \pi^+\pi^-$
produced by radiative return to the tail of the $\psi(2S)$ that can
pass the event selection criteria and yield fitted dilepton masses
around 3.097 GeV.  This is the main background to $\psi(3770)
\rightarrow J/\psi \pi^+\pi^-$, 
as shown in Fig.~\ref{fitmassplot}(b). 
Here the histogram shows the dilepton mass distribution for 
$\psi(2S) \rightarrow J/\psi \pi^+\pi^-$ 
from our Monte Carlo simulation.  The higher peak is due
to the radiative return to the $\psi(2S)$ peak, and the lower peak is from
radiative return to the tail of the  $\psi(2S)$.
The points
with error bars show the total contribution from $\psi(2S)$ and
$\psi(3770)$ production and decay.
From the simulation, we estimate that $6.0 \pm 0.5 \pm 0.6$
out of $17.8 \pm 4.8$ events in the peak near 3.1 GeV in
Fig.~\ref{fitmassplot}(a) are 
due to $\psi(2S) \rightarrow J/\psi \pi^+\pi^-$,
where the first error is statistical and the second
is the systematic arising from the uncertainty in the $\psi(2S)$
resonance parameters.  The probability that the 17.8 events observed
are due to a fluctuation of the $6.0$ events is
$3.8\times 10^{-4}$.

After background subtraction,
$11.8 \pm 4.8$ signal events are remained.
The branching fraction
for the non$-D \overline D$ decay
$\psi(3770) \rightarrow J/\psi\pi^+\pi^-$ is measured to be
$$BF(\psi(3770) \rightarrow J/\psi\pi^+\pi^-) =
 (0.338 \pm 0.137 \pm 0.082) \%. $$
Using $\Gamma_{\rm tot}$ from the PDG~\cite{PDG02},
this branching fraction corresponds to a partial width of
$$\Gamma(\psi(3770) \rightarrow J/\psi\pi^+\pi^-) =
(80 \pm 32 \pm 21)~~{\rm {keV}}. $$
More detail may be found in 
Ref.~\cite{psipptojpsipipi1} and Ref.~\cite{psipptojpsipipi2}.
%

%


\end{document}